# Plasmonic Metamaterial With Coupling Induced Transparency

Shuang Zhang, Dentcho A Genov, Yuan Wang, Ming Liu and Xiang Zhang[*]

5130 Etcheverry Hall, Nanoscale Science and Engineering Center, University of California, Berkeley, California 94720-1740, USA

**Abstract**: A plasmonic "molecule" consisting of a radiative element coupled with a sub-radiant (dark) element is theoretically investigated. The plasmonic "molecule" shows close resemblance to the electromagnetically induced transparency (EIT) in an atomic system. Because of its sub-wavelength dimension, this EIT-like "molecule" can be used as building block to construct a "slow light" plasmonic metamaterial.

Localized surface plasmon (LSP) with a deep sub-wavelength dimension has received much attention due to the interesting physics and important applications such as FRET and enhanced Raman scatterings [1, 2]. The interaction between two or more plasmonic particles gives rise to interesting hybridization of the plasmonic modes, which can be utilized to tune the resonance frequency of the system [3, 4]. Recently, the merge of plasmonics and metamaterials areas has led to the achievement of magnetic and negative index metamaterials in the optical frequencies, which may open up new perspective towards achieving ultimate control of light in the nano-scale dimension [5, 6, 7, 8].

Generally, a plasmonic mode can be either super-radiant (radiative mode) or sub-radiant (dark mode) depending on how strong an incident light from free space can be coupled into the plasmonic mode [9]. The radiative mode has a large scattering cross section and a low quality factor due to the radiation coupling. On the contrary, the dark mode normally has a significantly larger quality factor, which is only limited by the loss of the metal. Naturally, this leads to the analogy of the dark mode with a meta-stable level in an atomic system. The meta-stable energy level is necessary for the realization of electromagnetically induced transparency (EIT) medium [10, 11, 12]. In an atomic EIT system, the coupling between one energy level and a meta-stable level by a pump light leads to a destructive interference between different pathways and thus results in a narrow transparent window within a broader absorption band. Recently, an analog of EIT system using coupled micro-size optical resonators has been proposed [13], this, as well as some other related recent works [14, 15, 16], brings the original quantum phenomena into the realm of classical optics. In order to achieve the ultimate goal of nano-sized active optical circuit with switching and modulation capabilities, miniscule "light slowing" elements with deep sub-wavelength features are highly desired, while the dimensions of the current all-optical EIT systems are much larger than the wavelength. In this work, we propose a coupled radiative-dark plasmonic complex to mimic the functionality of an atomic EIT system, which might have the potential to be integrated with the nanophotonic circuits due to its ultra-small dimensions. In addition, we propose the first EIT-like optical metamaterial consisting of coupled nanoscale plasmonic resonators. The metamaterial has well defined effective properties that clearly show resemblance to the EIT in the atomic systems. This could enable interesting applications, such as slow light and enhanced nonlinear effects [17, 18].

---

[*] Corresponding author. Email: xiang@berkeley.edu

We begin by introducing a general decription of an EIT-like plasmonic "molecule" without specific configurations. The "molecule" consists of two artificial "atoms", a radiative plasmonic element which strongly couples with the light in the free space and a dark plasmonic element which weakly couples to the incident light, both have a resonant frequency $\omega_0$. The coupled system can be mathematically described in the linear regime as,

$$\begin{pmatrix} \delta + j\gamma_A & \alpha \\ \alpha & \delta + j\gamma_B \end{pmatrix} \begin{pmatrix} A \\ B \end{pmatrix} = \begin{pmatrix} gE_0 \\ 0 \end{pmatrix} \qquad (1)$$

where $A$ and $B$ are the resonance magnitudes of the radiative and dark "atoms", $\delta = \omega - \omega_0$ is the detuning of the frequency from the resonance, $\gamma_A$ and $\gamma_B$ are the damping factor with $\gamma_A \gg \gamma_B$, $\alpha$ is the coupling between the two "atoms", $g$ is a geometric parameter proportional to the scattering cross section of the radiative "atom", and $E_0$ is the electric field of the incident light. In Equation (1), we have assumed an ideal case, in which the dark plasmonic "atom" has a zero cross section, such that there is no coupling with the incident light at all. The amplitude of the dipole response of the radiative "atom" is given as,

$$A = \frac{gE_0(\delta + j\gamma_B)}{(\delta + j\gamma_A)(\delta + j\gamma_B) - \alpha^2} \qquad (2)$$

Short inspection of the above expression shows that the polarizability of the "molecule", which is proportional to the amplitude $A$ of the radiative "atom", has a very similar form as that in the EIT atomic system, especially in the case of zero detuning of the coupling laser frequency from the corresponding atomic transition [19]. Since the plasmonic "molecule" is normally deep subwavelength, an effective medium can be constructed by these building blocks. The susceptibility of the effective medium is then proportional to the amplitude $A$. Assuming a small detuning from the resonance frequency ($\delta \ll \gamma_A, \gamma_B, \alpha$), to the second order approximation, we can write

$$\chi = \chi_r + j\chi_i \propto \frac{\gamma_B^2 + \alpha^2}{(\gamma_A\gamma_B + \alpha^2)^2}\delta + j[\frac{\gamma_B}{\gamma_A\gamma_B + \alpha^2} + O(\delta^2)] \qquad (3)$$

The real part of susceptibility disappears at transparency frequency (zero detuning). Since the second order derivative of $\chi_r$ is zero, there is no dispersion in group velocity and a pulse centered at the transparency frequency will travel through the medium without distortion. In the ideal case of $\gamma_B = 0$, the imaginary part of susceptibility also disappears, leading to a totally transparent effective medium. From equation (3), the slope of the real parts of susceptibility is inversely proportional to $\alpha^2$, e. g. a smaller coupling coefficient will result in a slower group velocity. On the other hand, in the realistic case of a finite $\gamma_B$, $\alpha^2$ needs to be sufficiently large to keep $\chi_i$ close to zero. Thus, there is a tradeoff between a small loss and a large dispersion, which needs to be carefully balanced in the design of the EIT-like plasmonic elements.

Now we turn to a specific design of nano-plasmonic "molecules" for the realization the EIT-like system. Silver is selected to build the "molecule" because of its low loss. The plasma frequency of silver is $\omega_p = 1.366 \times 10^{16} s^{-1}$ and the damping rate $\gamma = 3.07 \times 10^{13} s^{-1}$, which are obtained by fitting the experimental data [20] with Drude model. It is well known that a simple metal strip functions as an effective optical dipole antenna in the optical region [21]. Such dipole antenna structure is thus chosen to serve as the radiative "atom" in the EIT-like plasmonic system. The resonance frequency of the metal strip can be readily tuned by varying its length. For a dipole antenna with a resonance in the visible range, the length can be significantly shorter than half of the wavelength in the free space since the coupling of light with the free electrons in the metal confines waves at the surface and reduces the effective wavelength [21]. Fig. 1(a) shows the amplitude of electric field versus frequency near one end of the strip, with the dimensions shown in the inset. The resonance frequency is around 429 THz ($\lambda = 699$ nm, more than 5 times the antenna length), with a quality factor ~ 11.8 estimated from the line-width of the resonance peak. The quality factor is mainly limited by the radiation dissipation, which comprises a large part of the total loss.

The dark "atom" consists of two parallel metal strips with a small separation, as shown in Fig. 1(b). This configuration has symmetric and anti-symmetric modes, whose resonances are separated in the frequency domain. The anti-symmetric mode has counter propagating currents on the two strips, therefore, there is no direct electrical dipole coupling with the radiation wave and it can be considered as the dark mode with significantly longer dephasing time. We note that this parallel-metal-strip configuration has served as an artificial magnetic dipole in the optical frequencies [7]. The anti-symmetric resonance frequency of the dark "atom" is closely aligned with that of the radiative "atom" [see Fig. 1(b)], as required by the general model described before. The quality factor of the dark "atom" is around 82, which is about 1 order of magnitude larger than that of the radiative "atom".

In contrast to an atomic EIT system, where the coupling between two energy levels is realized with a pump beam, the coupling between the radiative and dark "atoms" in the plasmonic EIT system is determined by their spatial separation [Fig. 2(a)]. Fig. 2 (b) shows both real and imaginary parts of the electric field probed at the end of the radiative metal strip [red arrow in Fig. 2(a)] for different separations between the radiative and dark "atoms". The field is the sum of the incident light and the dipole response of the radiative "atom"; as a result, it has a linear relation with the dipole polarizability. Thus, by investigating the field response we can gain insight into the effective susceptibility of the system. For all separations, the coupling between the radiative and dark "atoms" leads to a narrow dip at the center of the broad peak for the imaginary part of electric field, which confirms the EIT-like destructive interference between the two pathways: direct dipole excitation of the radiative "atom" from the incident wave, and excitation of the dark "atom" (by the radiative "atom") coupling back to the radiative "atom". At a large separation of 100 nm, where the coupling is weak, the contrast of the dip is very small, which is due to the finite quality factor of the resonance in the dark "atom". With the increase of coupling (decreasing separation), the dip widens and becomes deeper, which is similar to the quantum EIT in an atomic system. At the frequency corresponding to the transparency (dip in the imaginary part of the field), the real part of the electric field shows a highly dispersive behavior, indicating that a light pulse travels at a significantly slower group velocity in a metamaterial consisting of this coupled plasmonic system. To visualize this destructive interference between the two pathways, we compare the 2D distribution of electric field at 428.4 THz for the radiative antenna uncoupled [Fig. 2(c), left] and coupled with the dark "atom" [Fig. 2(c), right]. Without coupling to the dark "atom", the radiative antenna is strongly excited by the incident plane wave with high electric field forming at its end facets. By placing the dark "atom" 40 nm from the radiative one, the electromagnetic field is coupled back and forth between the radiative and dark "atoms", leading to a destructive interference and a suppressed state in the radiative "atom" with much weaker electric field at its ends.

Next, we propose an optical EIT-like metamaterial with the coupled radiative-dark "molecule" as the building block [Fig. 3(a)]. In the following discussion, the separation between the dark and radiative "atoms" is fixed at 40 nm, with all other parameters kept the same as before. To assure a minimum coupling between the neighboring unit cells, the plasmonic "molecules" are arranged periodically with a spacing of 400 nm in the *x-y* plane, and a spacing of 200 nm in the *z* (propagation) direction. Fig. 3(b) shows the transmission for metamaterials consisting of one to four layers of unit cells along the propagation direction. In the figures, the transmission exhibits a peak within a broader absorption band, which is similar to the transmission usually observed in an atomic EIT system. With the increase of the unit cells along the propagation, the transparent peak decreases due to the existence of loss in the metal. The peak transmission versus the number of layers is shown in Fig. 4(a). A linear relation between the natural logarithm of peak transmission versus number of layer is observed, indicating that a single propagating mode dominates for light traversing through the metamaterial. From the slope in Fig. 4(a) we can calculate the imaginary part of susceptibility to be 0.104. The real parts of the susceptibility for all the different number of unit cells are derived from the transmission phase shifts and shown in Fig. 4(b). The calculated susceptibilities are almost the same regardless of the number of unit cells along propagation, suggesting that the structure functions as a true optical effective medium. At the transparent peak frequency, $\chi_r$ is very close to zero (-0.012), which is characteristics of a typical EIT system. Finally, the group refractive index is estimated from the slope of $\chi_r$ at the transparency frequency, which is about 10.2. Thus, we demonstrate a plasmonic EIT-like metamaterial operating in the visible frequency, with group velocity one order of magnitude slower than that in the free space with a bandwidth of 25 THz, while at the transparency peak, the refractive index of the metamaterial is almost unity ($n = 0.994 + 0.052i$).

Though we only focused on a specific EIT-like metamaterial in the optical frequency, the general principle can be readily extended to longer wavelength such as THz and microwave frequencies, where the dissipative loss in the metamaterial is lower. Recently, a dark mode split ring resonator (SRR) with very high quality factor and deep subwavelength size was demonstrated in the microwave [22], this, when combined with another suitably designed radiative antenna, could lead to a slow light meta-media in the longer wavelengths.

In summary, we have demonstrated a novel plasmonic metamaterial which closely mimics the EIT phenomena in an atomic system. Although the slow down factor (~10) is much lower than the EIT in an atomic system, however, it has the advantages such as room temperature operation, wide bandwidth and more importantly, ready integration with nanoplasmonic circuit. This plasmonic metamaterial, when incorporated with active or nonlinear dielectrics, may lead to dynamically tunable group velocities for light traveling through it. This would open up a new avenue towards the manipulation and control of light in very broad frequency range.

This work was supported by AFOSR MURI (Grant No. 50432), SINAM and NSEC under Grant No. DMI-0327077.

**Figure captions:**

Fig. 1: (color online) (a) Left: the schematic of a silver optical antenna which functions as a radiative "atom". The geometric parameters $W_1$, $L_1$ and $t$ are 50 nm, 128 nm and 20 nm, respectively. A plane wave is incident along $z$ direction, with $\vec{E}$ and $\vec{H}$ along $x$ and $y$ directions. Right: the spectral response of an $E_x$ probe (red arrow) placed 10 nm from the center of the end facet of the antenna. (b) Left: the schematic of a pair of metal strips forming a dark plasmonic "atom". The geometric parameters $W_2$, $L_2$ and $s$ are 30 nm, 100 nm and 30 nm, respectively. The incident wave vector lies in the y-z plane and is tilted $45°$ from the z direction, with electric field along x direction (perpendicular to the metal strips). Right: the spectral response of an $E_y$ probe (red arrow) placed 10 nm from the center of the end facet of one of the metal strips in the dark "atom".

Fig. 2: (color online) (a) Top view of the plasmonic system consisting of a radiative element and a dark element with a separation $d$, with light incident at normal direction (b) The real part and imaginary part of a $E_x$ probe placed at 10 nm from the end facet of the radiative antenna [red arrow in Fig. 2(a)] for separations ranging from 40 nm to 100 nm between the radiative and dark elements. (c) the 2D field plot of an uncoupled radiative "atom" (left) and a radiative "atom" coupled with a dark "atom" with a separation of 40 nm (right) at frequency of 428.4 THz, as indicated by the red triangle in Fig. 2(b).

Fig. 3: (a) schematic of a plasmonic EIT metamaterial (b) The transmission spectra for one, two, three and four layers along the propagation direction.

Fig. 4: (a) (color online) The natural logarithm of peak transmission at transparency frequency vs. the number of unit cells along propagation direction. (b) The real part of susceptibility vs frequency for 1 (gray), 2 (red), 3 (green) and 4 (blue) layers along propagation direction.

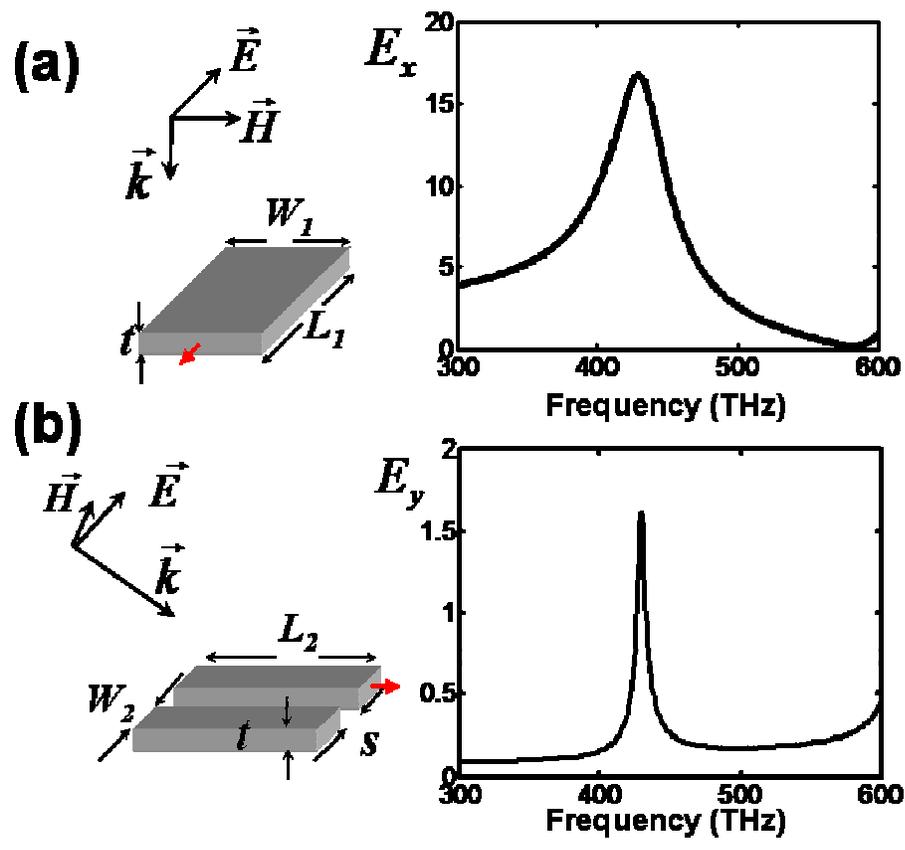

Fig. 1

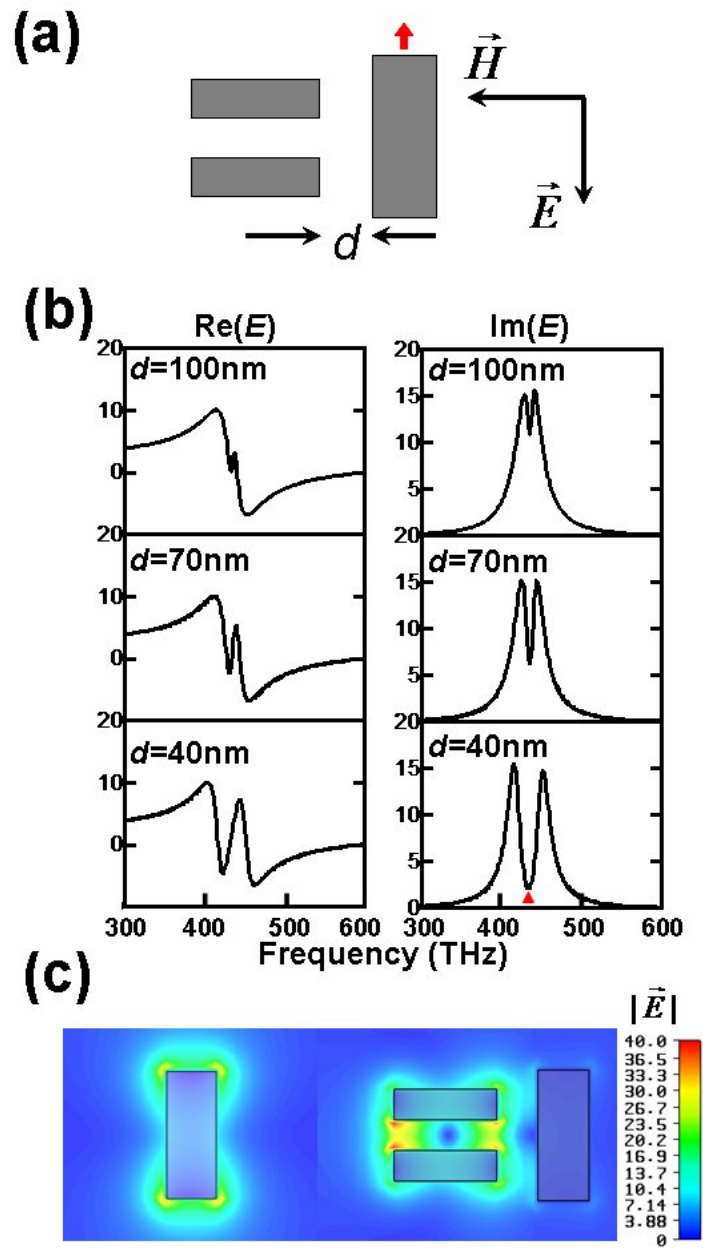

Fig. 2

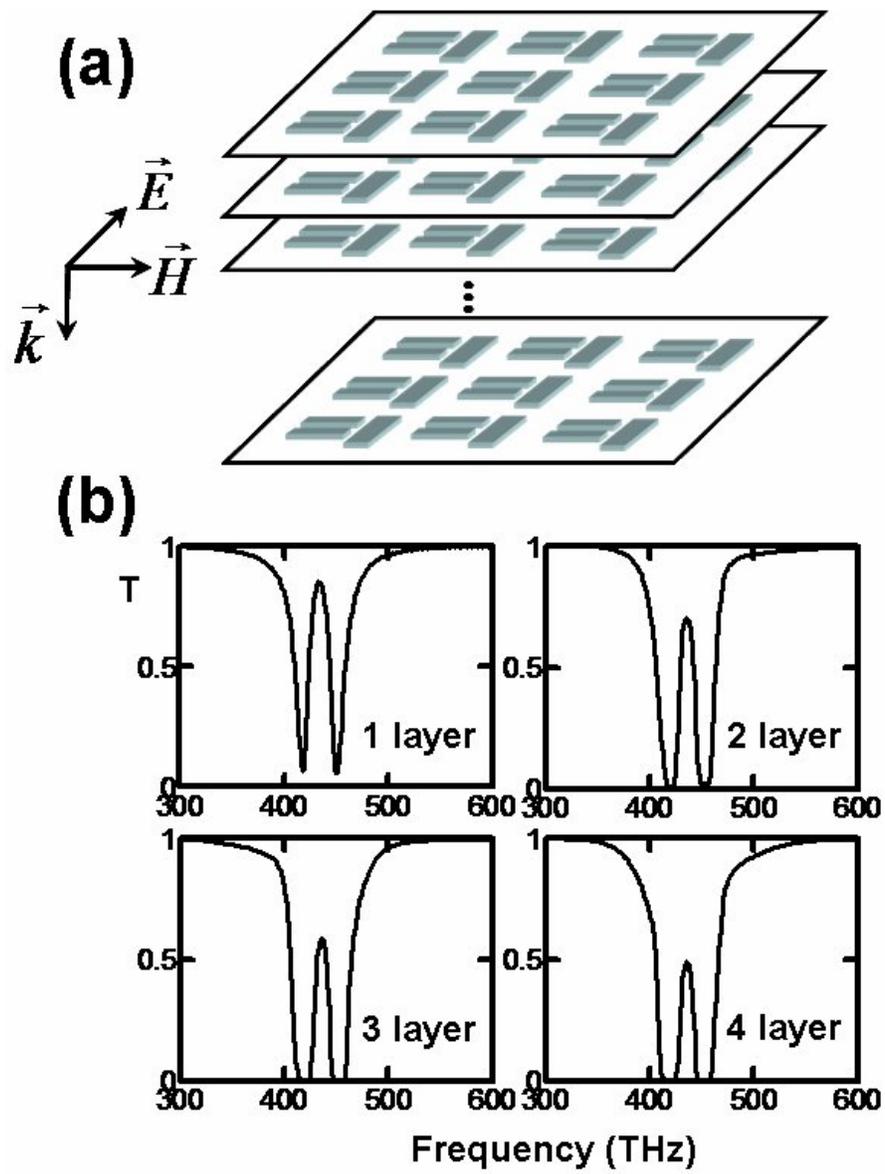

Fig. 3

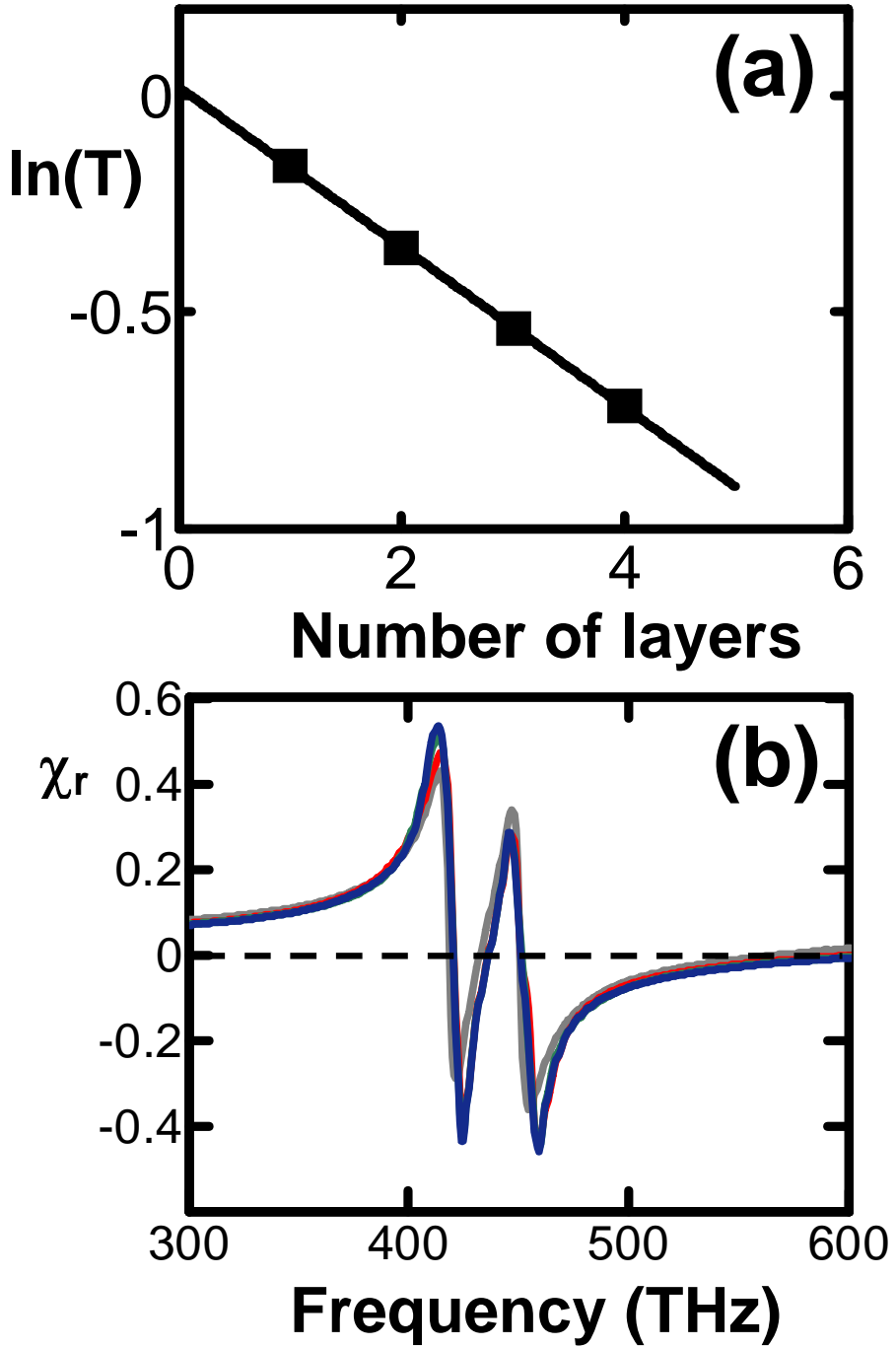

Fig. 4